# Self-Induced Buckling in Hollow Microgels

Leah Rank* and Emanuela Zaccarelli*

ACCESS | Metrics & More | Article Recommendations | Supporting Information

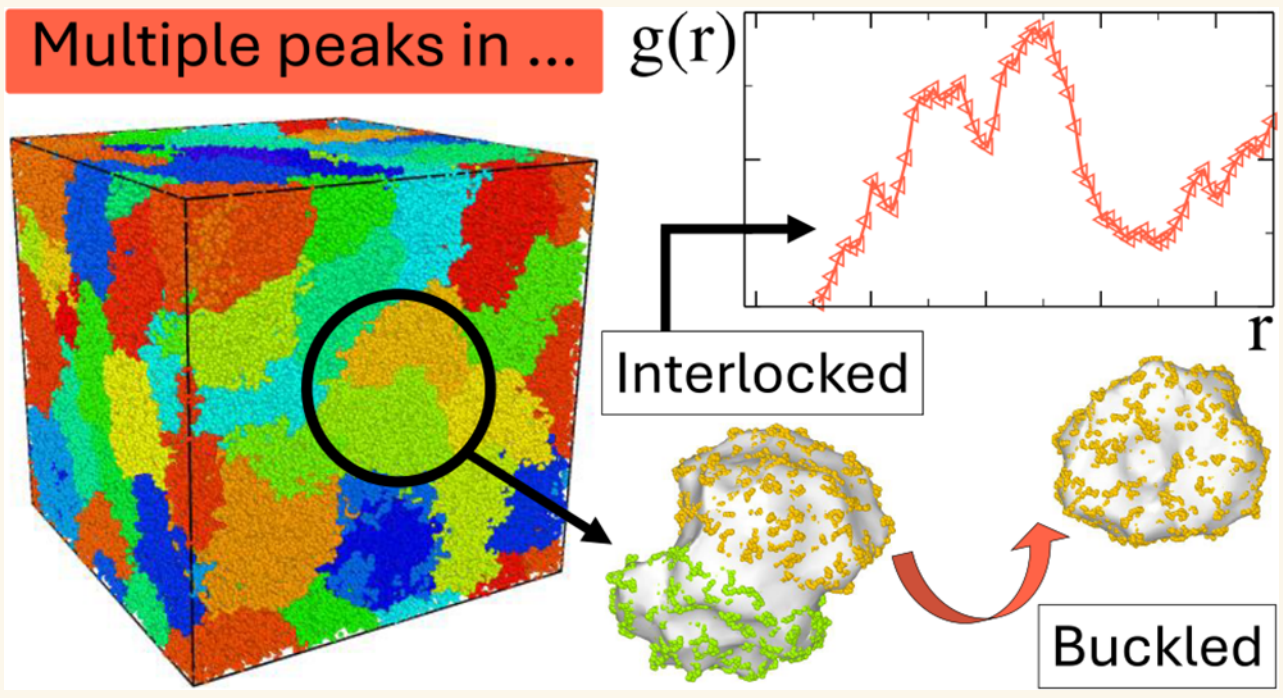

**ABSTRACT:** Hollow microgels are elastic polymer shells easily realizable in experiments. Recent works have shown the emergence of buckling events in dilute hollow microgels under the effect of an added osmotic pressure. Here, we perform large-scale simulations to show that these microgels at high enough packing fractions undergo spontaneous symmetry-breaking deformations ranging from single large dents to multiple indentations, even in the absence of any externally applied stress. This self-induced buckling phenomenon is thus solely driven by interparticle crowding. We construct a phase diagram inspired by vesicle shape theories, mapping local curvature metrics as a function of the reduced volume, to quantify these findings, and we also propose ways to observe the occurrence of buckling in experiments. The present results thus rationalize the deformations occurring in suspensions of micro- and nanoscale elastic shells, offering a synthetic analogue to biological ones and allowing direct control on buckling instabilities for potential applications. Beyond materials design, these insights may also help to describe shape regulation in natural systems such as cells and vesicles, where similar deformations are observed.

Colloidal suspensions have long captivated scientists because they offer a mesoscopic analogue to atomic and molecular systems.[1,2] Their larger particle sizes allow for direct observation with optical microscopy, while their interactions and collective behavior can be finely tuned. As a result, colloidal suspensions serve as model systems for studying fundamental processes such as crystallization, glass formation, and self-assembly—insights that are not only relevant for condensed matter physics but also for designing advanced materials. Hard-sphere colloids have played a central role in colloidal science, with packing fraction as their primary control parameter.[1] However, soft colloids exhibit even richer phase behavior due to their deformability and tunable interactions. This softness allows for crystal-to-crystal transitions, multiple glassy states, and complex jamming behaviors, often inaccessible to hard-sphere systems.[3−8]

Microgels, in particular, are highly versatile soft colloids composed of a cross-linked polymer network.[9] Their sizes typically range from tens of nanometers to several micrometers, and their defining feature is their ability to respond to external stimuli. Depending on their chemical composition, microgels can reversibly swell or deswell in response to changes in temperature,[10−12] pH,[13,14] or ionic strength.[15,16] In this respect, thermoresponsive microgels undergo a transition from a swollen state in a good solvent to a collapsed state when the temperature is raised above a certain value, due to the reduced affinity to the background fluid, typically water. One of the most widely studied systems is based on poly(*N*-isopropylacrylamide) (pNIPAM), which exhibits a volume phase transition (VPT) temperature around 32 °C.[17] The combination of tunable size, softness, and interparticle interaction makes thermoresponsive microgels ideal candidates for both fundamental research[18−20] and material design. They are widely employed in applications ranging from optical sensing[21] to drug delivery and rheological modifiers.[22−25] At the same time, their collective behavior continues to attract attention in soft matter physics, especially in the context of phase transitions, elasticity, and glassy dynamics.[18−20,26]

While different overall microgel topologies have been synthesized in recent years,[27−29] hollow microgels—spherical polymer network shells—appear especially promising. Not only is the inner cavity appealing for encapsulation[30] and controlled release[31] of a cargo, but from a fundamental point of view, hollow microgels exhibit striking similarities to biological systems such as vesicles and cells.[32,33] Studying their collective behavior and mechanical response thus provides an ideal model system for understanding phenomena related to shape

Received: September 16, 2025
Revised: December 24, 2025
Accepted: December 24, 2025

deformation in nature,[34−36] such as buckling. In this context, initially rather spherical soft shells, being either vesicles,[37,38] red blood cells[39−41] or hollow microgels,[32,42] have been experimentally observed to buckle, i.e., to collapse asymmetrically under stress into double-layered, bowl-like structures.

Being highly tunable experimentally, hollow microgels offer a synthetic model to explore this intriguing phenomenon in detail. In particular, the experimental studies so far have focused on the buckling of individual microgels induced by osmotic pressure through the addition of polymer chains. For practical purposes, it is, however, crucial to unveil whether these shape deformations may also occur in crowded conditions and without the addition of external agents. Indeed, the internal degrees of freedom of the polymer shells and their ability to highly shrink and deform may provide alternative ways to pack hollow microgels at high densities, utterly different from their nonhollow counterparts. To answer this question, we resort to state-of-the-art monomer-resolved computer simulations of realistic microgels, which allow us to gain microscopic knowledge of individual shape changes at the single microgel level even under extreme crowding. In this respect, it is worth mentioning a recent study of binary mixtures of hollow microgels and nonhollow microgels,[43] which has shown how the presence of hollow microgels suppresses the tendency of crystallization of the regular ones due to their larger deformability.

While computational studies have explored the bulk behavior of nonhollow microgels in detail,[20,44] a similar investigation for a suspension of hollow microgels is still lacking. To perform such a study, we previously characterized single thermoresponsive hollow microgels[45] to determine the ideal characteristics, i.e., amount of cross-linkers in the network and shell thickness, needed to maintain the presence of a stable cavity close and above the VPT temperature. Building on this, we now perform extensive Molecular Dynamics simulations of an ensemble of hollow microgels in good solvent, i.e., at low temperature, and systematically vary microgel concentration up to well above random close packing. We find an intriguing behavior, not present in nonhollow microgels, where microgels strongly deform and exhibit fascinating shapes. While previous studies[32] have observed buckling in microgels using external stress caused by added polymer chains, the present work demonstrates a spontaneous route to microgel buckling in a pure system of hollow microgels, induced by the mutual microgel interaction. This instability emerges at random positions in bulk suspensions at high enough densities as a result of a competition between packing and elasticity. We therefore define this phenomenon as 'self-induced buckling', to distinguish it from externally induced buckling. While both mechanisms share an increase of pressure over microgels, causing the occurrence of buckling, there could be distinctive features of the self-induced case that we aim to unveil with the present work. We thus rationalize these findings in the context of geometric observables, used to quantify shape deformation in vesicles, providing evidence of the optimal region of parameters where this self-buckling phenomenon should be found experimentally.

Due to the relevance of hollow microgels as a proxy for biological systems, the present work provides a framework to understand how to induce shape deformations within packed states of elastic shells and how to gain fundamental control over this phenomenology for enhancing their potential for applications.

## RESULTS AND DISCUSSION

### Structure of Hollow Microgels with Increasing Concentration: Single-Particle Properties

We start by analyzing the single-particle properties of a suspension of hollow microgels with increasing packing fraction. The latter is quantified as the nominal packing fraction $\zeta$, defined in the Methods section by eq 10, as usually done for soft colloids.[46] We focus on microgels with a relative shell thickness $\delta_{rel}$ = 0.275 and two different values of the cross-linker concentration $c$ = 5% and $c$ = 10%. The former case closely describes the experimental system studied in ref 42., as shown in our previous work.[45]

We report the behavior of the average size of the microgels, quantified either by the radius of gyration $R_g$ or by the hydrodynamic radius $R_H$, as a function of $\zeta$ in Figure 1. All radii are normalized by their respective values in the dilute limit ($\zeta$ = 0) to favor the comparison between the two studied values of $c$ and the corresponding data for regular (nonhollow) microgels in ref 20. We find that all the data tend to a similar decay at large $\zeta$, roughly compatible with isotropic shrinking $\sim\zeta^{-1/3}$ (dashed line in Figure 1). However, important differences are found in the behavior of $R_g$ and $R_H$ for hollow microgels, as compared to nonhollow ones. The radius of gyration is found to be independent of $c$, and its normalized value is always lower than that for regular microgels. Instead, the hydrodynamic radius shows an opposite trend: it decreases considerably less than in the nonhollow case, and it further increases with increasing $c$. This indicates that $R_g$ in hollow microgels essentially probes the cavity size, which does not depend on $c$ for the same $\delta_{rel}$ and is more compressible than the dense core of regular microgels. Instead, the variation of $R_H$ is mostly determined by the shell, whose elasticity increases with $c$ and exceeds that of the fuzzy corona of standard microgels, allowing it to resist more effectively to shrinking. Interestingly, in the case of the asymmetric binary mixtures studied in ref 43, the shrinking of the hollow microgels quantified by a similar quantity to $R_H$ was found to be much more pronounced than for the nonhollow ones. This may be due to the fact that hollow microgels shrink more easily with respect to the regular ones also present in the mixture, as opposed to the pure hollow case. Indeed, the situation is quite different for the present suspension of hollow microgels, where the overall shrinking is significantly reduced.

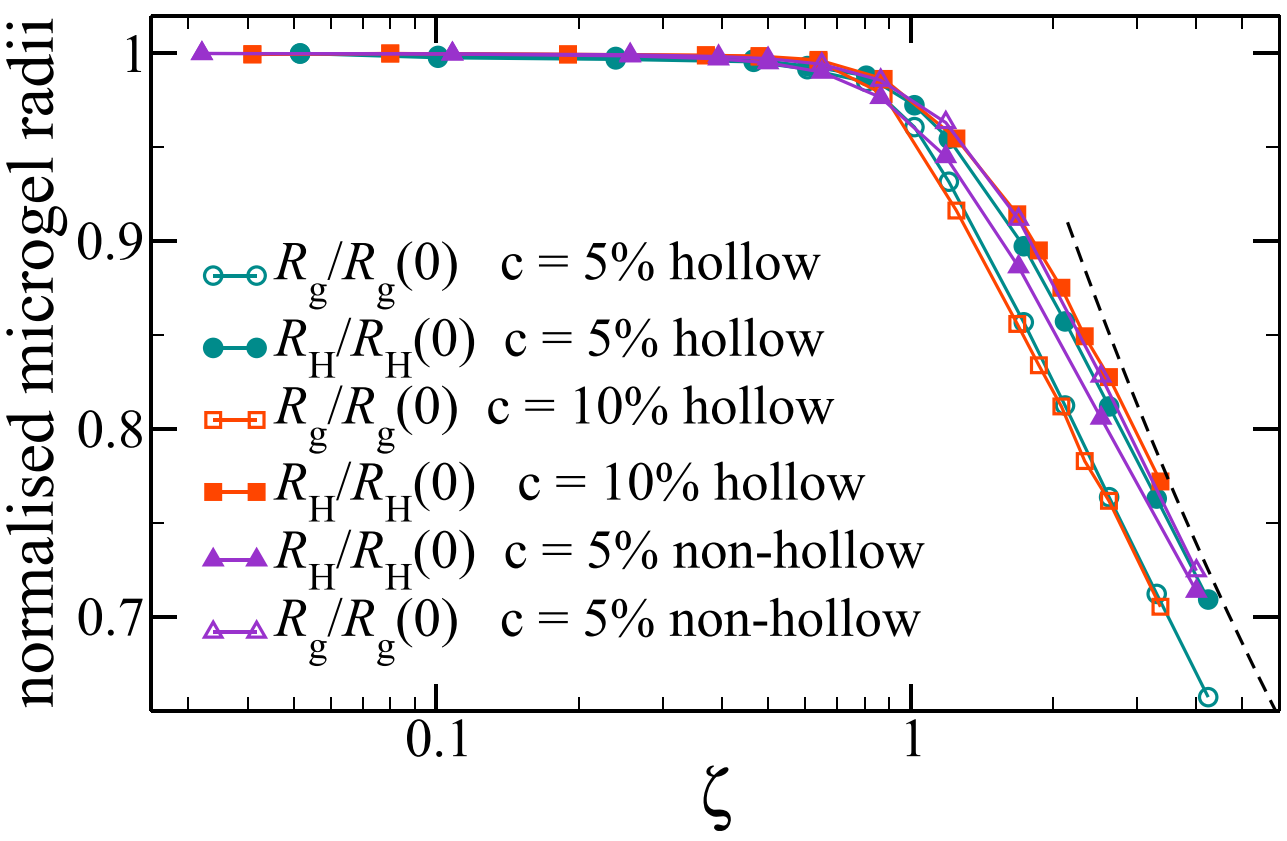

**Figure 1.** Size of the microgel versus nominal packing fraction $\zeta$. The figure reports both the normalized radius of gyration $R_g/R_g(0)$ and the normalized hydrodynamics radius $R_H/R_H(0)$, where the reference values are those of single microgels ($\zeta$ = 0), and compares hollow microgels of initial shell thickness $\delta_{rel}$ = 0.275 and $c$ = 5% and $c$ = 10% with nonhollow ones with $c$ = 5% studied in ref 20. The dashed line indicates isotropic shrinking, i.e., a $\zeta^{-1/3}$ behavior.

However, when we go beyond the average values but actually look at the full distributions of the radii, important changes between the two cross-linker concentrations emerge. Figure 2 reports $p(R_g)$ at intermediate and high packing fractions for $c$ = 5% and $c$ = 10% hollow microgels in panels (a) and (b), respectively. The softer microgels are found to continuously shrink with increasing $\zeta$, always maintaining a Gaussian-like $R_g$-profile, similar to nonhollow microgels with the same $c$,[20] as shown in the Supporting Information (SI), see Figure S1. However, for the latter, as $\zeta$ increases, the distributions get significantly higher and narrower, while for the hollow microgels, they tend to stay similar to each other, with some narrowing only at very large $\zeta$. A very different picture manifests for the microgels with a higher degree of cross-linking, whose distributions actually get smaller in height and clearly display strong deviations from a Gaussian distribution above $\zeta \sim 1$. In particular, for $\zeta \gtrsim 1.7$, two distinct peaks are visible in the distribution, as emphasized in the inset of Figure 2b. Both peaks can be individually described with a Gaussian distribution, clearly suggesting that there are two populations of slightly larger and slightly smaller microgels. The two peaks persist upon further increasing $\zeta$, and their presence is found to depend on the preparation protocol. Indeed, at these very high densities, the microgels are essentially glassy, *i.e.*, as shown by their mean-squared displacements reported in the SI (Figure S2). In this respect, it is therefore important to discuss how the final density is reached. We remark that for the present data, each packing fraction is obtained from the previous one in an annealed sequence, while results from a sudden quench would just shrink the microgels much more rapidly, without giving them the time to adapt and thus suppressing fluctuations. This issue is further discussed in the SI (see Figures S3 and S4).

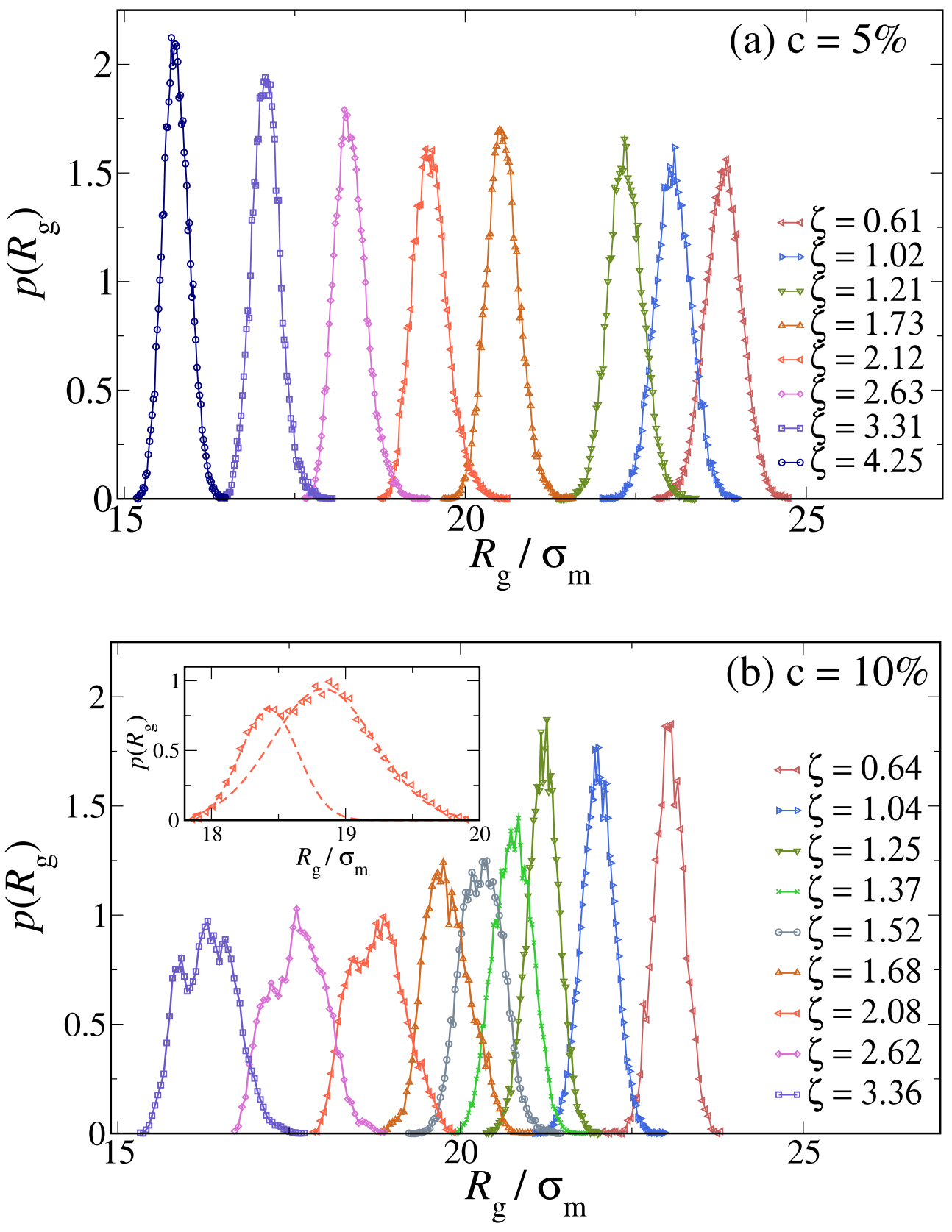

**Figure 2.** Radius of gyration distribution $p(R_g)$ of hollow microgels with $c$ = 5% (a) and $c$ = 10% (b). The inset in (b) shows $p(R_g)$ at $\zeta$ = 2.08 where two peaks are clearly visible, which can be individually fitted with two separate Gaussians (dashed lines).

The emergence of non-Gaussianity in the distribution of $R_g$ for $c$ = 10% indicates an important shape change in at least some of the microgels, despite the average $R_g$ continuing to decrease as in all other cases, including regular microgels. To investigate this aspect further, we also calculate the distributions of asphericity $a$, defined in the Methods section by eq 9, for individual microgels at each studied packing fraction. The resulting $p(a)$ are reported at selected values of $\zeta$, similar for the two types of microgels, in Figure 3. While in dilute conditions, the shape distributions of the microgels are rather similar and close to that of a sphere, more and more asymmetry develops with increasing $\zeta$, which is much more pronounced for $c$ = 10%. In particular, for low cross-linking, the hollow microgels always retain a single-peak distribution, although this becomes rather broad, i.e., with a significant increase of its variance, at high $\zeta$. This broadening is greatly enhanced at high cross-linking, where the distribution first becomes rather flat, finally developing the onset of different particle populations at very high $\zeta$, as signaled by the emergence of almost two distinct peaks. It is again important to note that while $R_g$, reported above, essentially probes the cavity size, the asphericity, which is calculated via the convex hull enclosing the whole microgel, quantifies an overall variation of the shape of the particles.

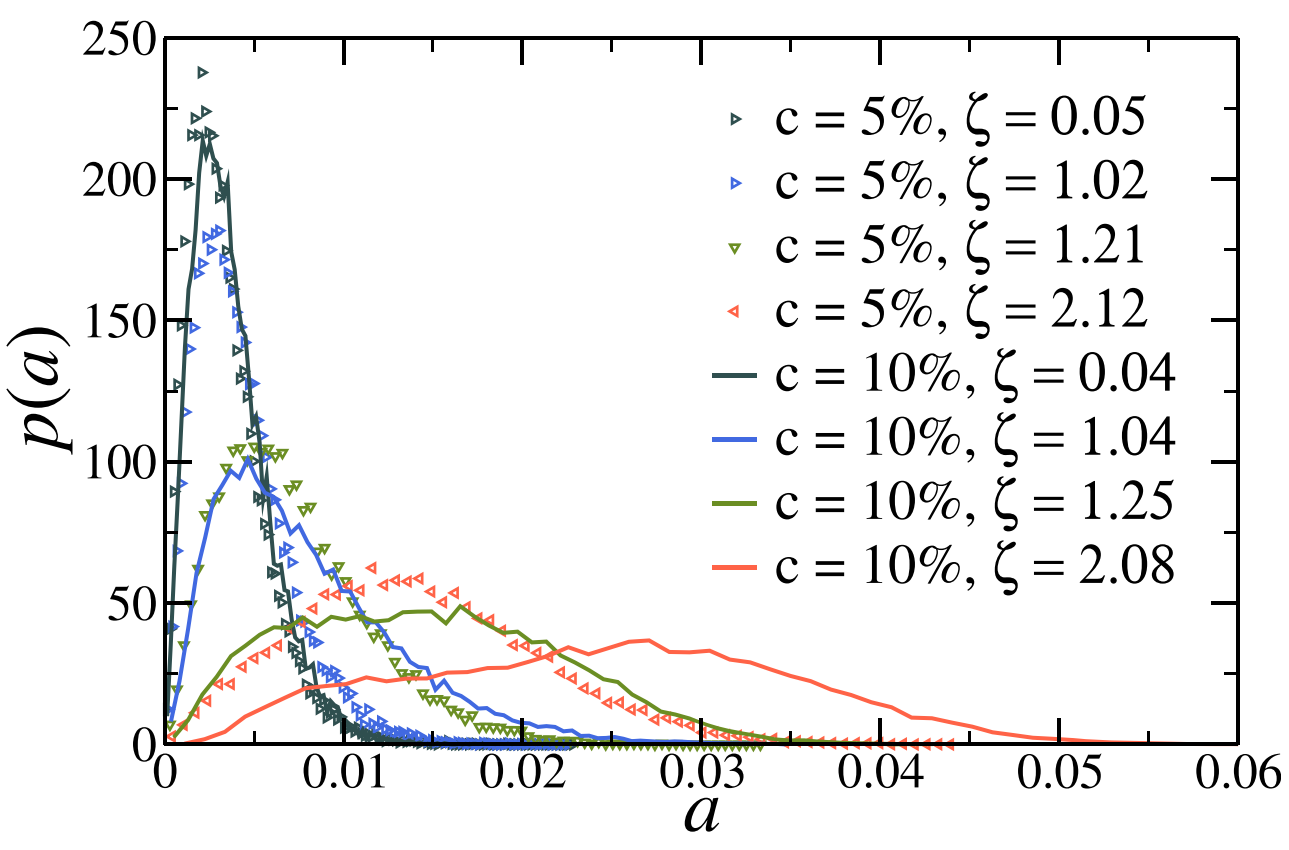

**Figure 3.** Asphericity distribution $p(a)$ of hollow microgels with $c$ = 5% (symbols) and $c$ = 10% (solid lines) at selected corresponding packing fractions (highlighted by the same color coding).

The qualitative difference in both observables $p(R_g)$ and $p(a)$ between the two cross-linker regimes is also directly visible in the snapshots reported in Figure 4. Although the two systems are displayed at a similar packing fraction, they show a very different behavior, where the softer microgels in Figure 4a are organized in a looser and more spherical manner with respect to their more elastic and therefore tighter counterparts, showing fascinating deformations highlighted in (b).

## Structure of Hollow Microgels with Increasing Concentration: Collective Properties and Buckling

Having looked at individual particle properties, we now try to assess the implications of these modifications on the whole suspension. To this aim, the obvious quantity to monitor is the radial distribution function $g(r)$ of the microgels' center of mass, which gives us information on the local structure of the system.

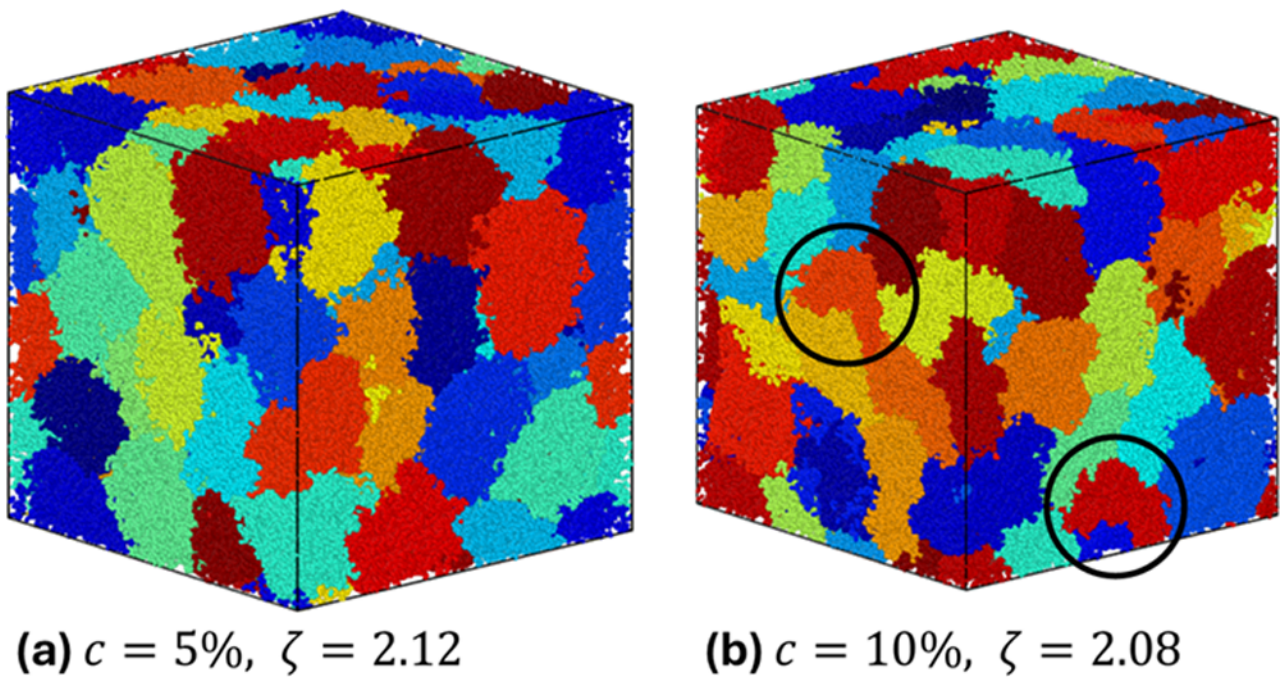

**Figure 4.** Snapshots of the simulated system for the two studied microgels: $c = 5\%$ (a) and $c = 10\%$ (b) at a comparable packing fraction $\zeta \sim 2.1$. The black circles highlight buckled examples with one big dent.

In Figure 5, we report $g(r)$, where the radial distance $r$ has been rescaled by the microgel diameter $\sigma_m$, for hollow microgels with $c = 5\%$ (a) and $c = 10\%$ (b) at different packing fractions, comparable between the two cross-linker regimes. Starting from dilute conditions, the peak position moves to smaller and smaller distances as $\zeta$ increases, while the height of the peak increases. This holds up to slightly above random close packing, namely $\zeta \sim 0.8$. Above this packing fraction, the peak position of the $g(r)$ continues to decrease but its amplitude also decreases with further increasing $\zeta$, in a so-called reentrant behavior. These features are in agreement with what was observed for regular microgels for $c = 5\%$.[20] However, this behavior only persists up to a certain packing fraction for hollow microgels. At even higher $\zeta$, additional peaks develop in stark contrast to regular microgels. In particular, the appearance of multiple peaks is already visible at moderate packing fraction, $\zeta \sim 1.0$, for $c = 10\%$, later developing a cascade of small peaks at larger and larger $\zeta$. In particular, the multiple peaks are well visible at $\zeta \sim 2$ for both studied hollow microgels, being more evident for $c = 10\%$ where also two distinct main peaks are present, both decorated with multipeaks. We argue that this is a genuine feature of concentrated suspensions of hollow microgels, since the presence of the multiple peaks is not due to statistical noise, but to the fact that selected preferred distances arise, caused by mutual indentation of hollow microgels, which does not arise in regular microgels.

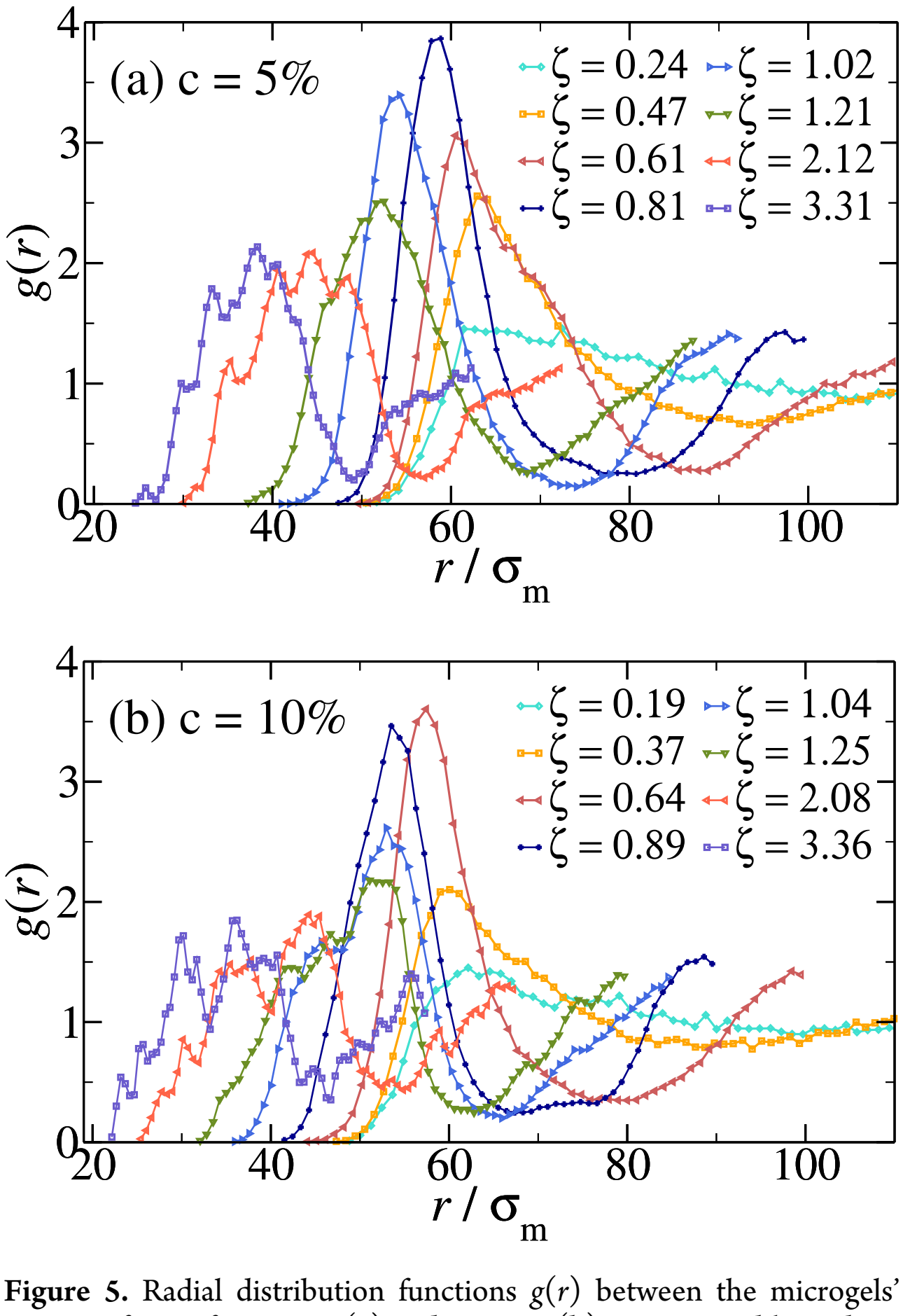

**Figure 5.** Radial distribution functions $g(r)$ between the microgels' centers of mass for $c = 5\%$ (a) and $c = 10\%$ (b), at comparable packing fractions. Both plots have been rescaled on the $x$-axis by the monomer diameter $\sigma_m$.

To highlight this peculiar feature arising in suspensions of hollow microgels, we take a close look at $c = 10\%$ for a few selected packing fractions and identify the occurrence of these peaks between nearest neighbors at the specific distances.

We thus focus on $g(r)$ and on its multiple peaks for $c = 10\%$ hollow microgels, plotted again in Figure 6e for chosen values of $\zeta$. We start by looking at $\zeta = 1.25$ where a first additional peak occurs at $r/\sigma_m \sim 40$, which is labeled as (a) in Figure 6e. This corresponds to the occurrence of moderate dents in the microgels, as illustrated in the snapshot (a), where a (green) microgel surrounded by its nearest neighbors is highlighted, together with the (dark gray) neighbor found at the distance that belongs to the smallest peak of $g(r)$. Looking more closely, it appears evident that all the transparent microgels are found at a distance belonging to the main peak in $g(r)$ representing simple faceting, while the dark gray neighbor is found at a closer distance, which induces the presence of a dent. This dent causes the microgel to slowly fill up its cavity, evident when comparing its density profile $\rho(r)$ in Figure 6 f (green curve) to the averaged profile in dilute suspensions $<\rho(r)>$ (gray curve). Next, we move on to examine a larger packing fraction, i.e., $\zeta = 2.08$, where multiple clear peaks occur, labeled as (b, 2), (b, 1), and (c)—in correspondence with the snapshots. In this state point, we find, for example, the red microgel, surrounded by three neighbors (shown with a colored surface mesh and white particles) within a distance contributing to peak (b, 2) with its light gray neighbors and to peak (b, 1) with the dark gray one. The microgel is deformed to a large extent, and each of these nearest neighbors induces a dent; three of them are visible in the snapshot (there is also an additional one present in the back). Going down in distance and focusing on peak (c), the smallest for the packing fraction under study, the snapshots show deeply dented interlocked microgels that have completely lost their sphericity. These largely deformed states, where the cavity is completely absent as shown by the density profiles in Figure 6f, can be defined as buckled. Interestingly, the first appearance of the peaks seems to occur in coincidence with the packing fraction where the cavity starts to be filled ($\zeta = 1.25$ in Figure 6f and also in the average density profiles reported in Figure S5), and this is the mechanism eventually leading to self-induced buckling.

Finally, at the highest investigated packing fraction $\zeta = 3.36$, we notice—focusing on the same microgel as in snapshot (b)—that it only shrinks while preserving the same identical dents, as it is clear from the snapshot reported in panel (d). Overall, the whole $g(r)$ for $\zeta = 3.36$ appears to be shifted to the left with respect to the one for $\zeta = 2.08$, indicating just a shrinking of the system as a whole. The reason for this is that, at very high densities, the microgels are completely trapped by their neighbors, so that they maintain their deformed shape as is. This regime of final shrinking at fixed deformation was also detected in regular microgels at very high $\zeta$[20,47] and is consistent with the $\zeta^{-1/3}$ behavior of the microgel size, as shown in Figure 1. It is also worth noting that, although the dents do not change position, their depth overall decreases due to the shrinking of the microgels, so that the buckling becomes less pronounced once exceeding a certain packing fraction and is best observed at intermediate $\zeta$ values. This is not inferred simply from the evolution of $g(r)$ over $\zeta$ but becomes evident when comparing the red microgel snapshot in Figure 6b to the same microgel in violet in panel (d) at a lower packing fraction, which displays less deep dents after additional shrinking than it did in (b).

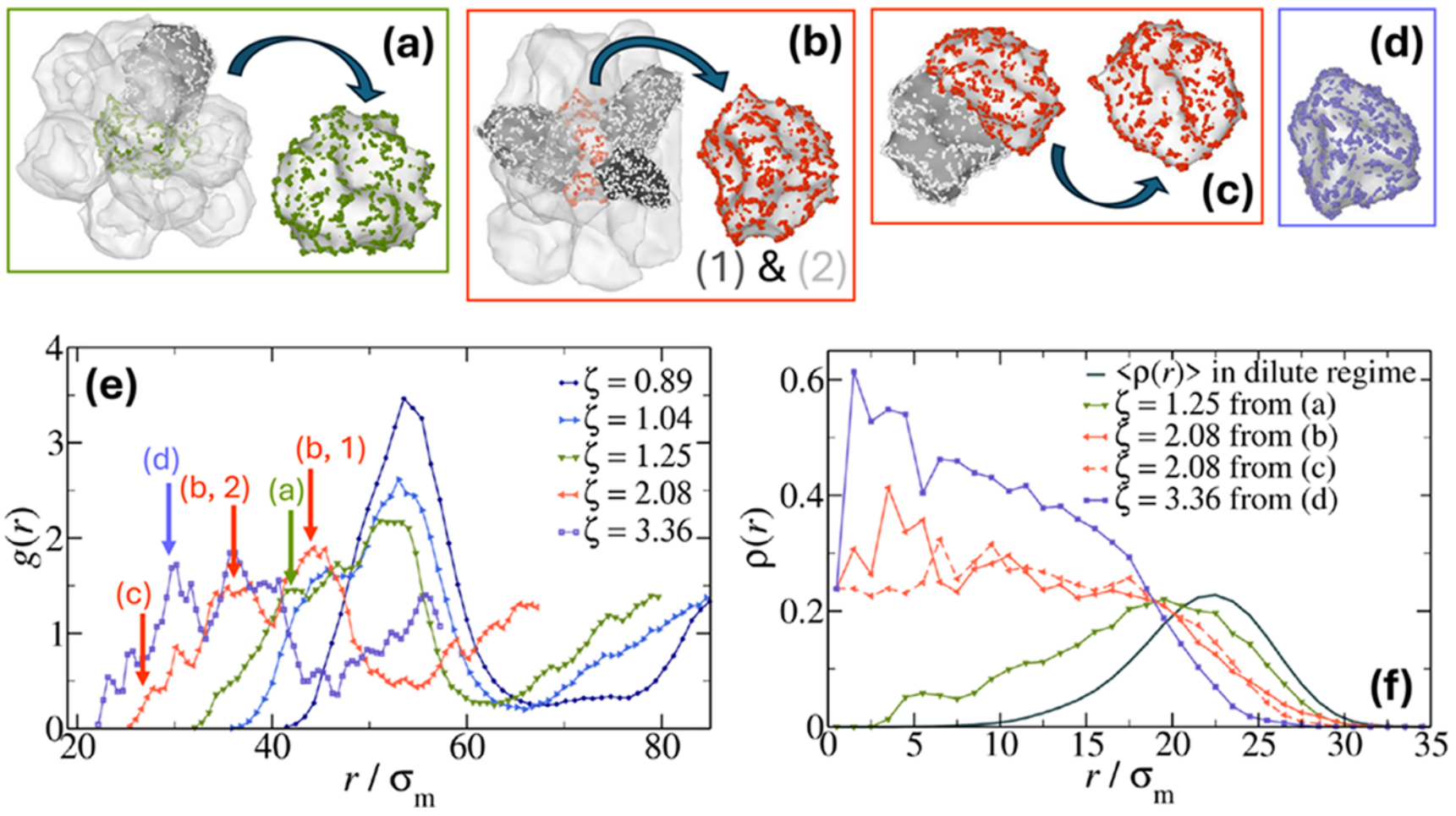

**Figure 6.** Snapshots of nearest-neighbor hollow microgels with $c$ = 10% distanced corresponding to $g(r)$ peaks plotted in Figure 5. The peaks are highlighted by color-coded arrows in panel (e) using a color-consistent map indicating packing fractions. The central microgel is highlighted separately, following the same color scheme, in each of the plots as (a) a green microgel at distance $\sim 42\ \sigma_m$ from the microgel with white particles and a gray surface mesh for $\zeta = 1.25$; (b) a red microgel distanced $\sim 44\ \sigma_m$ (labeled as 1) from the microgel with the dark gray surface mesh and $\sim 35\ \sigma_m$ (labeled as 2) from the two gray ones for $\zeta = 2.08$; (c) another red microgel at distance $\sim 28\ \sigma_m$ with respect to the gray microgel again for $\zeta = 2.08$; (d) the same microgel as in (b) but now in violet and at $\zeta = 3.36$, showing that at these very high densities no further deformation occurs; (e) radial distribution functions of the suspension as in Figure 5b, highlighting the peaks at which the microgels from the previous snapshots are found; and (f) density profiles of the microgels highlighted in (a–d), compared to the dilute limit (full curve without symbols).

The occurrence of several dents, each made by a different neighbor, is thus responsible for the multiple peaks in the $g(r)$, appearing to be a distinctive feature of the elastic deformability of hollow microgels that should be clearly visible in confocal microscopy experiments.[32,42] To provide robustness of these results, we also report the behavior of $g(r)$ with increasing $\zeta$ for another system of hollow microgels, still characterized by $c$ = 10%, but with a smaller thickness $\delta_{rel} = 0.21$, in the SI (see Figure S6). Also, in this case, multiple peaks arise at high enough packing fractions. These thinner microgels were observed to display a largely anisotropic shape also at the individual level in response to increasing temperature.[45] However, we prefer to focus on the $\delta_{rel} = 0.275$ case reported throughout the manuscript, because its shell thickness is comparable to that of recent experiments carried out by Hazra and co-workers.[42]

## Shape Phase Diagram

Having detected buckling events in crowded hollow microgel suspensions, we now try to quantify their shapes, inspired by studies on vesicles[48−51]—objects that share many similarities but also important differences with the present hollow microgels. We thus build a *shape phase diagram* reporting on the $x$-axis the microgel's reduced volume $v$ and on the $y$-axis the integrated and normalized mean curvature $\Delta a$. While all details on the calculation are provided in Methods, we report here the basic definitions of these observables. In brief, the reduced volume is defined as,

$$v = \left(\frac{R_V}{R_A}\right)^3 \tag{1}$$

where the radii $R_V$ and $R_A$ are obtained from the volume $V$ and the surface area $A$, respectively, of the surface mesh enclosing the microgels, i.e.,

$$R_V = \left(\frac{3V}{4\pi}\right)^{1/3}, \quad R_A = \left(\frac{A}{4\pi}\right)^{1/2} \tag{2}$$

The second quantity $\Delta a$ is related to the bilayer-coupling model used for vesicles,[38,52,53]

$$\Delta a = \Delta A/(8\pi R_A d) \tag{3}$$

where $d$ is the shell thickness of the hollow microgel and $\Delta A$ is defined as,

$$\Delta A = \mathrm{d}\oint C_1 + C_2 \mathrm{d}A = \mathrm{d}\oint 2H\mathrm{d}A \tag{4}$$

with $C_{1,2}$ the principal curvatures measured at every triangle face of the surface mesh and $H$ the corresponding mean curvature (see Methods for more details on the calculation).

The shape phase diagram is $(v, \Delta a)$ reported in Figure 7 for both studied cross-linker concentrations, where each point represents one microgel at a given state point. Individual results are reported in order to appreciate the large fluctuations from microgel to microgel. In particular, panel (a) shows results for $c$ = 5% at all packing fractions examined, while the corresponding results for $c$ = 10% are shown in panel (b). In both cases, $\Delta a$ decreases with increasing $\zeta$, which might stem from the reduction in bumpiness of the surface due to the overall shrinkage in microgel size. However, a striking difference between the two types of microgels is observed for the variation of the reduced volume: while for the softer microgels, the reduced volume is essentially independent of $\zeta$, a clear variation of $v$ emerges in the stiffer case. This becomes even more evident by looking at the averaged shape phase diagrams in Figure 8, where the results for each packing fraction are averaged over all microgels as well as over different time steps. Focusing on the low cross-linker case in (a), the average reduced volume $<v>$ is found to only slightly shift to smaller values for intermediate packing fractions, suggesting very weak deformations, before perhaps increasing again toward more spherical shapes at the largest studied $\zeta$. Within error bars, we can consider the variation of $<v>$ for these microgels to be absent. Instead, the reduction in $<\Delta a>$ becomes statistically significant above $\zeta \sim 1.0$, while it is not present below random close packing. A very different situation presents itself for the high-cross-linked microgels, shown in Figure 8b. Essentially, no variation in either $<\Delta a>$ or $<v>$ is observed for low $\zeta$. However, for $\zeta \sim 1.0$, a sudden and robust decrease of $<v>$ is detected, while $<\Delta a>$ still remains roughly constant. This happens within a very narrow region of $\zeta$, corresponding to the start of the buckling region of the phase diagram. At larger values of $\zeta$, $<\Delta a>$ starts to decrease with $v$ changing more moderately. Interestingly, at very large $\zeta$, also in this case, it seems that $< v >$ starts to reincrease. This last regime, arising for both types of microgels, is observed when they have reached maximum deformation.

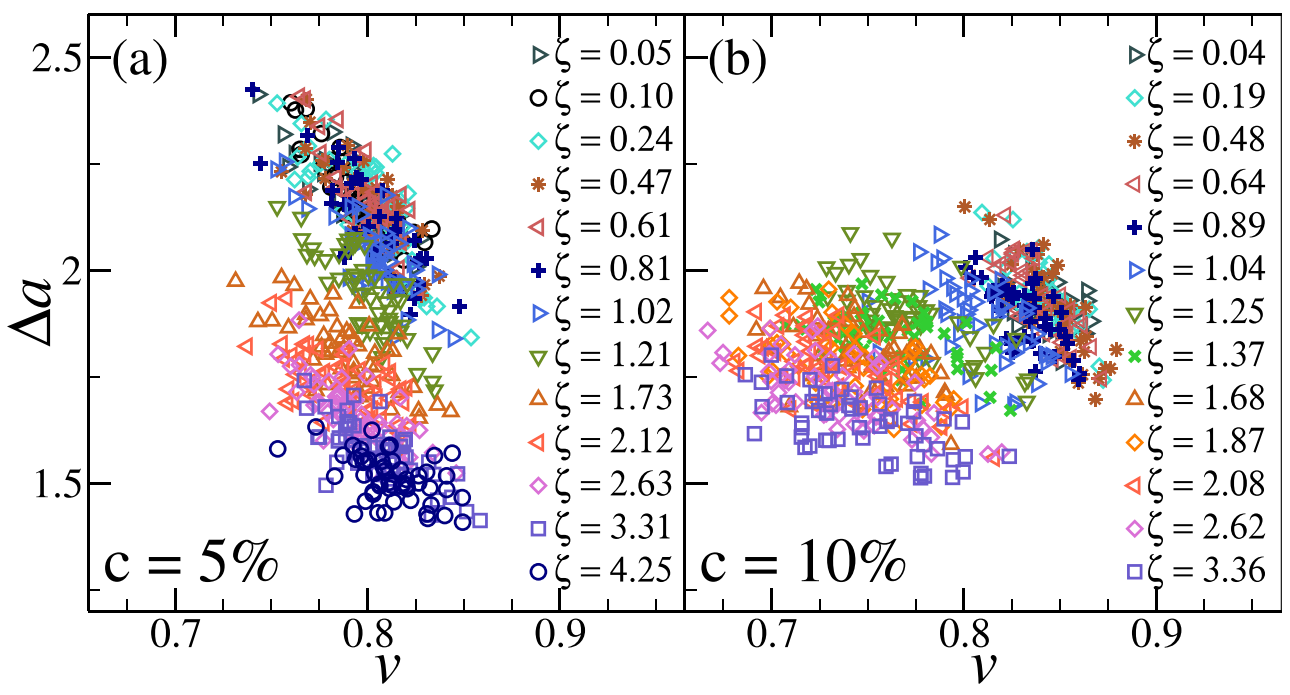

**Figure 7.** Shape phase diagram, representing the reduced volume $v$ versus the normalized mean curvature $\Delta a$, for hollow microgels with (a) $c$ = 5% and (b) $c$ = 10% at different packing fractions. Each symbol represents an individual microgel, and all calculations are performed on the final equilibrated configuration. All surface meshes have been created with $s = 30$, with $r_P = 4.5\ \sigma_m$ used in (a) and $r_P = 4.0\ \sigma_m$ in (b), as discussed in Methods and the SI.

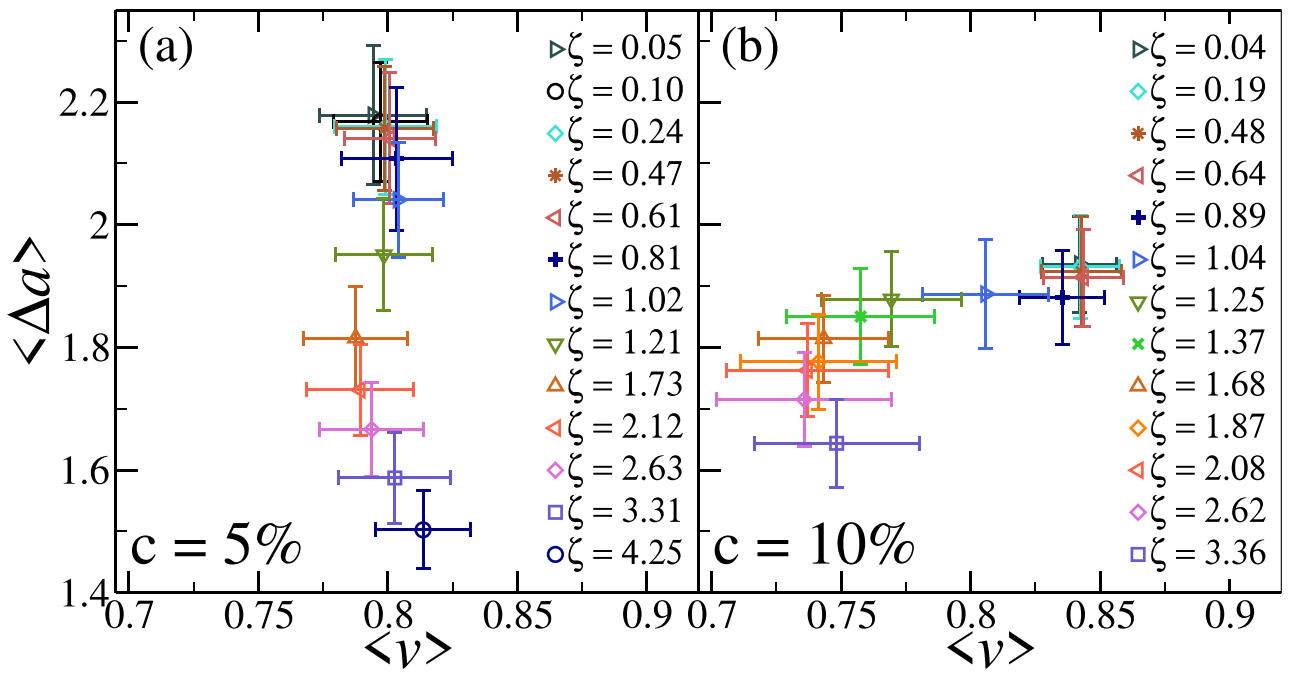

**Figure 8.** Averaged shape phase diagrams. Same as in Figure 7 but now data are ensemble-averaged over all microgels and time-averaged over five different configurations. Standard deviations are plotted as error bars.

Importantly, the two averaged shape phase diagrams in Figure 8 are reported in the same ($<v>$ , $<\Delta a>$) region and show a much more moderate variation of $<\Delta a>$ for $c$ = 10% microgels accompanied by a more significant change of $<v>$.

To interpret these findings, we recall that we adopt two complementary shape descriptors and, by construction, a perfect sphere has $v = \Delta a = 1$. The reduced volume $v$ measures the deviation of an object from the most compact shape with a given surface area, and thus captures the degree of "deflation" or internal compaction, while $\Delta a$ captures the degree of surface bending through the integrated mean curvature. In our simulations, we observe two regimes—see Figure 8a vs b. For softer microgels ($c$ = 5%), crowding mainly leads to an isotropic compression: the particles shrink quite uniformly, but they do not undergo large-scale shape deformations. Accordingly, the average reduced volume remains almost unchanged. By contrast, more elastic microgels ($c$ = 10%) respond to crowding differently: rather than simply shrinking, they develop large-scale dents and shape anisotropies at intermediate packing fractions. In this case, the reduced volume decreases significantly, reflecting the fact that the overall shape departs from the initially fairly spherical one in dilute systems. When it comes to $\Delta a$, however, we do not observe a clear distinction between the two cases. The overall local bumpiness of the microgel surface, which is slightly smoother for $c$ = 10% leading to smaller values of $\Delta a$, seems to overpower any curvature signal from a global shape change. Hence, in both cases, $\Delta a$ simply decreases with packing fraction since the microgel surfaces become less bumpy, the more densely the microgels are packed, and therefore have to shrink into themselves.

Finally, we notice that analogous calculations performed for $c$ = 10% hollow microgels with smaller thickness, i.e., $\delta_{rel} = 0.21$, as reported in the SI (Figure S7), are qualitatively similar to the findings obtained for the larger value of the thickness, yielding generality to our findings for thin enough hollow microgels.

## CONCLUSIONS

In this work, we investigated via molecular dynamics simulations the behavior of hollow microgels in suspension at different packing conditions. In particular, we compared two types of microgels differing in cross-linker concentration $c$ and thereby in their elastic properties. Microgels were deliberately chosen to be relatively thin shells ($\delta_{rel} = 0.275$), a value used in recent experiments,[42] so that at high enough elasticity ($c$ = 10%), they are prone to deformation, as observed in single-particle behavior.[45] Here, we have extensively assessed the role of crowding on the deformation of the microgels upon further increasing the packing fraction up to very dense states. The early manifestation of the role of elasticity is observed in the behavior of the single-particle properties, such as the distribution of the gyration radii. While for softer microgels, $p(R_g)$ retains a Gaussian shape, similar to the nonhollow case, strong deviations appear for more elastic ones, which clearly display the onset of different microgel populations at intermediate values of $\zeta$, which then persist up to the largest studied densities. This is also evident in the asphericity distribution, again revealing multiple populations for the more elastic microgels, some of which undergo very large deformations, not reached for the softer case. It is important to reiterate that this deformation, being a self-induced behavior by nominally all identical microgels, is entirely attributable to shape fluctuations, arising from the competition between crowding and elasticity. Therefore, there is no sharp buckling transition of the whole suspension, and only some of the microgels, experiencing the largest shape fluctuations, undergo the local buckling phenomenon. A more collective response may be achievable by the application of an external force, such as shear, which should be able to drive a collective buckling, which will be investigated in the future.

However, also under these *spontaneous* conditions, it is possible to notice the occurrence of the buckling of individual microgels in a sea of less deformed ones. This is visible in the radial distribution function of the suspension, which at high enough $\zeta$, shows the clear evidence of multiple peaks, not observed in nonhollow microgels,[20] but which could be experimentally detected by confocal microscopy. This peculiar feature arises due to the variety of microgel shapes occurring in the whole suspension, manifesting in numerous small peaks in $g(r)$, that are clearly attributable to the deformation of the microgels having multiple dents, as shown in snapshots of Figure 6. Importantly, the experimental morphologies of buckled microgels under external pressure observed in ref 32. seem to be more compatible with a bowl-like shape with a single dent, so that multidented conformations could be a distinctive feature of the present self-induced buckling by neighbor microgels.

Since in the present simulations each microgel is different from the others, due to their internal degrees of freedom, we resorted to metrics usually employed for vesicles in order to properly capture their shape fluctuations. To this aim, we analyzed their surface mesh at a quantitative level by plotting the shape parameter $\Delta a$ related to the overall curvature of the microgels against its reduced volume $v$ in what we call a shape phase diagram, where the individual microgel fluctuations are very evident, particularly close to the onset of buckling. While softer microgels mostly deswell and experience faceting in response to crowding, the more elastic case shows a clear transition at a packing fraction of ∼1.0. This is evidenced by a clear discontinuity in reduced volume observed in the shape phase diagram of Figure 8b for $c$ = 10% hollow microgels and signals the onset of buckling in some of the particles. Indeed, looking at the averaged density profiles shown in the SI (Figure S5), the transition happens roughly where the microgels start to close their cavity. In this respect, this effect is the counterpart of what is observed for a single microgel by raising the temperature,[45] where the competition between elasticity and shrinking makes the microgel lose its cavity. This cavity loss in crowded environments could thus also be detectable in experiments, by measuring the form factors under crowded conditions using selective deuteration as previously done for nonhollow microgels.[54]

It is very instructive to compare the shape phase diagrams obtained for our hollow microgels with respect to vesicles. For the latter, having a smoother surface, a value close to the spherical state, characterized by $v = \Delta a = 1$, is usually reached in their most relaxed form. Instead, for hollow microgels, the normalized mean curvature seems to be a less meaningful parameter, due to their bumpy surfaces. Still, the decreasing trend for the reduced volume $v$, which is caused by increasing deformation, i.e., number and depth of dents, takes reasonable values, aligning with what is observed in vesicle studies.[38,52,53] Indeed, it is this parameter that signals the onset of the buckling instability. It would be important in the future to try to measure this parameter also from confocal microscopy experiments[32] and to try to bridge the conceptual gap between hollow microgels and vesicles, to eventually find a theoretical description connecting the microgel shapes with their elasticity properties, similar to theories used to describe vesicle shapes - such as the relaxed model, bilayer-couple models,[51,52] or the area-difference-elasticity model[50] - and at least qualitatively predict the shape variations of hollow microgels.

It is also important to stress that at high packing fraction, the microgels are almost arrested (see Figure S2). While the study of the dynamics and the occurrence of a glass transition requires a separate dedicated study due to the large number of microgels (and of their monomers) and the needed long simulation times, we anticipate that it is of fundamental interest to inspect the interplay of buckling with possible nonmonotonicity of dynamical quantities, as predicted for Hertzian potentials[55] but not yet observed in nonhollow microgels.[20] Likewise, it will also be appealing to perform simulations or experiments under shear to amplify the buckling transition in the whole suspension and to investigate the response of these strongly deformed states in order to classify them within the wide panorama of soft colloids with intriguing rheological behavior.[56]

Finally, the present observations of dented and faceted morphologies draw direct analogies to the deformations seen in cells, vesicles, and other membrane-bound structures under confinement or crowding. Since the microscopic parameters of the hollow microgels are well-defined and tunable, our model system provides a unique opportunity to disentangle the mechanical principles underlying these shape transitions. This, in turn, may help to gain crucial insights into the complex responses of natural soft compartments, offering a bridge between synthetic soft matter and biological physics.

## METHODS

### Simulation Details

The numerical assembly protocol for disordered polymer networks is based on earlier works[57] using the oxDNA simulation package,[58] adapted to obtain hollow microgels[45,59] within a spherical shell. The network contains $N_c = cN_m$ cross-linkers, where $c$ is the molar fraction of cross-linkers, usually referred to as cross-linker concentration, and $N_m$ is the total number of monomers. We fix $c$ = 5 or 10% in the two sets of simulations considered, while the shell thickness $\delta_{rel} = (Z_{out} - Z_{in})/Z_{out}$ = 0.275, with $Z_{in}$ and $Z_{out}$, the inner and outer radii of the initial shell during assembly. The value of $\delta_{rel}$ is well within the experimental range and has been selected to observe the buckling phenomenon, based on results from our previous study.[45]

All the monomers and cross-linkers in the network interact by means of the bead−spring model of Grest and Kremer,[60] which mimics polymeric interactions. Namely, all particles experience an excluded volume interaction, modeled by the Weeks−Chandler−Andersen (WCA) potential:

$$V_{WCA}(r) = \begin{cases} 4\epsilon\left(\left(\frac{\sigma_m}{r}\right)^{12} - \left(\frac{\sigma_m}{r}\right)^{6}\right) + \epsilon & \text{if } r \leq 2^{1/6}\sigma_m \\ 0 & \text{otherwise} \end{cases} \quad (5)$$

In addition, particles that are chemically linked are also subjected to a finite extensible nonlinear elastic (FENE) potential, which prevents bond rupture and maintains the structural integrity of the network:

$$V_{FENE}(r) = -\epsilon k_F R_0^2 \ln\left(1 - \left(\frac{r}{R_0\sigma_m}\right)^2\right), \quad \text{if } r < R_0\sigma_m \quad (6)$$

The parameter $\epsilon$ sets the unit of energy, while $\sigma_m$, the monomer diameter, is the unit of length. All particles have unit mass $m_m$ and the spring constant of the FENE potential is set to $k_F$ = 15 with the maximum bond extension being $R_0$ = 1.5.

In order to be able to simulate a significant number of microgels, we set $N_m \sim 13000$, for which the network is large enough to be able to sustain its cavity, as shown in ref 45. We thus replicate the initially assembled microgel to conduct simulations of an ensemble of them. Specifically, we perform NVT Molecular Dynamics simulations of $N_{mgel} = 54$ microgels at a fixed temperature, $T^* = k_B T = 1.0$, where $k_B$ is the Boltzmann constant in a cubic box of length $l_{box}$ with periodic boundary conditions. To keep the temperature constant, we employ the stochastic velocity rescaling thermostat[61] following a leapfrog integration scheme with a time step $\delta t^* = \delta t\sqrt{\epsilon/(m_m \sigma_m^2)} = 0.002$. To properly capture the behavior of the system, we start from a very dilute regime of microgels in a large box, which gets gradually reduced with enough relaxation time ($5 \times 10^6$ timesteps) to allow the system to equilibrate at each stage. We thus perform simulations for an additional $2 \times 10^7$ timesteps in a wide range of packing fractions $\zeta$ from the dilute limit up to very concentrated conditions ($\zeta > 3.0$). For selected state points, we also compare a fast change in density to this slow annealing process, as discussed in the SI.

### Calculated Observables

We quantify the microgel size by calculating its radius of gyration, defined as

$$R_g = \sqrt{\frac{1}{N_m}\sum_{i}^{N_m}(\vec{r}_i - \vec{r}_{cm})^2} \tag{7}$$

where the vector $\vec{r}_i$ refers to the position of the $i$-th monomer and $\vec{r}_{cm}$ to the microgel's center of mass.[62] We also monitor the hydrodynamic radius $R_H$, that we calculate as in previous works:[63,64]

$$R_H = \left\langle 2\left[\int_0^\infty \frac{1}{\sqrt{(a_1^2+\theta)(a_2^2+\theta)(a_3^2+\theta)}}\mathrm{d}\theta\right]^{-1}\right\rangle \tag{8}$$

Here, the quantities $a_1$, $a_2$, $a_3$ are the principal semiaxes of the gyration tensor of the convex hull enclosing the microgel, which are also used to evaluate the microgel's asphericity:[2]

$$a = \frac{3(a_1^2 + a_2^2 + a_3^2)}{2(a_1 + a_2 + a_3)^2} - \frac{1}{2} \tag{9}$$

From the hydrodynamic radius of a single microgel, i.e., in dilute conditions, $R_{H,dilute}$, we determine the nominal packing fraction $\zeta$, that is the quantity usually monitored in experiments, here defined as,

$$\zeta = \frac{4\pi R_{H,dilute}^3 N_{mgel}}{3 l_{box}^3} \tag{10}$$

By varying $l_{box}$, we thus change $\zeta$, which can greatly exceed close packing, since microgels can shrink and deform upon increasing concentration.

The internal structure of the microgels is monitored by calculating their individual density profiles, $\rho(r)$ as a function of the distance $r = |\vec{r}|$ from the microgel center of mass, defined as

$$\rho(r) = \left\langle \sum_{i=1}^{N_m} \delta(|\vec{r} - \vec{r}_i|) \right\rangle \tag{11}$$

where angled brackets represent an average over several equilibrated configurations at different timesteps.

### Calculation of the Shape Phase Diagram

In order to make a connection with buckling phenomena observed in thin elastic shells and in vesicles, we calculate two observables that compose what we call a "shape phase diagram".[38,48−53] In particular, on the $x$-axis, this reports the microgel's reduced volume, which is defined as in eq 1. The radii $R_V$ and $R_A$ are calculated by constructing the surface mesh around the simulated microgel using the $\alpha$-shape method implemented in the OVITO software.[65,66] This method constructs a three-dimensional, triangulated surface by performing a Delaunay tessellation of the microgel coordinates, applying a probe sphere of radius $r_P$. The radius $r_P$ determines the resolution of the resulting mesh: smaller values yield a tighter fit that captures finer structural details, while larger values produce a smoother, more convex surface. Additionally, the mesh can be controlled by a smoothing parameter $s$. How those input quantities affect our results is illustrated in the SI. We choose to work with parameters $r_P = 4.5\,\sigma_m$ for $c = 5\%$, $r_P = 4.0\,\sigma_m$ for $c = 10\%$, and $s = 30$ for both cross-linker cases, which provide qualitatively robust and meaningful results, as extensively discussed in the SI and reported in Figures S8−S10. From these calculations, we get the vertices and faces of this triangular mesh around the microgel and can thus compute its volume $V$ as well as its surface area $A$, from which we then calculate $R_V$ and $R_A$ (see eq 2).

On the $y$-axis of the shape phase diagram, we plot the second reduced quantity related to the bilayer-coupling model used for vesicles,[38,52,53] $\Delta a$ defined in eqs 3 and 4. It can be observed that the thickness $d$ of the membrane cancels out when calculating $\Delta a$.

The principal curvatures $C_{1,2}$ are measured at every triangle face of the surface mesh. To compute the mean curvature $H$ at each vertex of the triangulated surface mesh, we use a discrete approximation based on the cotangent-weighted Laplace-Beltrami operator. For each triangular face, we compute the cotangent weights corresponding to its internal angles and accumulate them into a sparse Laplacian matrix $L$. At the same time, we estimate the local surface area around each vertex using the barycentric (Voronoi) area of the surrounding faces. The mean curvature vector at each vertex $i$ is then given by

$$\mathbf{H}_i = \frac{1}{2A_i}\sum_j L_{ij}(\mathbf{x}_j - \mathbf{x}_i) \tag{12}$$

where $\mathbf{x}_i$ is the position of vertex $i$, $L_{ij}$ are the cotangent weights, and $A_i$ is the local area associated with vertex $i$. The scalar mean curvature $H_i$ is obtained by taking the norm of $\mathbf{H}_i$. Since the curvature contribution is ultimately integrated over the surface, we compute the mean curvature per triangle face by averaging the scalar curvatures of its three vertices. The integrated mean curvature $\Delta A$ is then approximated by summing the face-averaged values weighted by the corresponding triangle areas:

$$\Delta A \approx \sum_f 2H_f A_f \tag{13}$$

where $H_f$ and $A_f$ are the mean curvature and area of face $f$, respectively. If we perform these calculations around a perfectly spherical particle, $H = 1/R$ is constant everywhere on the sphere and related to its radius $R$, eventually leading to $v = \Delta a = 1$.

## ■ ASSOCIATED CONTENT

### Data Availability Statement

Data for this article will be made available via ZENODO.

### Supporting Information

The Supporting Information is available free of charge at https://pubs.acs.org/doi/10.1021/acsnano.5c15990.

Additional analyses and discussion, including distributions of gyration radii for nonhollow microgels; mean-squared displacement of the hollow microgels; results of quenching protocols; averaged density profiles of the microgels and identification of cavity closure; additional results for hollow microgels with a thinner shell; and justification of the chosen Ovito parameters for surface mesh construction (PDF)

## ■ AUTHOR INFORMATION

### Corresponding Authors

**Leah Rank** − *CNR Institute of Complex Systems, Uos Sapienza, 00185 Roma, Italy; Department of Physics, Sapienza*

*University of Rome, 00185 Roma, Italy;* orcid.org/0009-0005-3411-1184; Email: leah.rank@uniroma1.it

**Emanuela Zaccarelli** − *CNR Institute of Complex Systems, Uos Sapienza, 00185 Roma, Italy; Department of Physics, Sapienza University of Rome, 00185 Roma, Italy;* orcid.org/0000-0003-0032-8906; Email: emanuela.zaccarelli@cnr.it

Complete contact information is available at: https://pubs.acs.org/10.1021/acsnano.5c15990

### Author Contributions

Author contributions are defined based on CRediT (Contributor Roles Taxonomy). Conceptualization: E.Z.; formal analysis: L.R., E.Z.; funding acquisition: E.Z.; investigation: L.R., E.Z.; methodology: L.R., E.Z.; project administration: E.Z.; supervision: E.Z.; validation: L.R., E.Z.; visualization: L.R.; writing—original draft: L.R., E.Z.; and writing—review and editing: L.R., E.Z.

### Notes

The authors declare no competing financial interest.

## ■ ACKNOWLEDGMENTS

The authors thank Primoz Ziherl and Giovanni Del Monte for useful discussions. The authors are also grateful to the CINECA award, under the ISCRA initiative, for the availability of high-performance computing resources and support. They acknowledge financial support from the European Union (HorizonMSCA−Doctoral Networks) through the project QLUSTER (HORIZON-MSCA-2021-DN-01-GA101072964). E.Z. also acknowledges support from ICSC—Centro Nazionale di Ricerca in High-Performance Computing, Big Data and Quantum Computing-(Grant No. CN00000013, CUP J93C22000540006, PNRR Investimento M4.C2.1.4).

## ■ REFERENCES